\documentclass[twocolumn,10pt]{revtex4}
\graphicspath{{./pic/}}

\def\degr{\ensuremath{^\circ}}
\newcommand{\ind}[1]{_{\text{\scriptsize #1}}}
\newcommand{\partder}[2]{\dfrac{\partial #1}{\partial #2}}
\newcommand{\grad}{\mathop{\mathrm{grad}}\nolimits} 
\newenvironment{pict}%
	{\begin{figure}[t]\begin{center}\noindent}{\end{center}\end{figure}}
\newenvironment{widepict}%
	{\begin{figure*}\begin{center}\noindent}{\end{center}\end{figure*}}
\def\look#1{(see Fig.~\ref{#1})}

\def\ms{\ensuremath{\,\text{ms}^{-1}}}

\begin{document}
\title{Analysis of Thermal Conditions of the 6-m BTA Telescope Elements
and the Telescope Dome Space}
\author{E.\,V.~Emelianov}
\affiliation{Special Astrophysical Observatory}
\begin{abstract}
The results obtained using the temperature monitoring systems of the 6-m BTA telescope
primary mirror, dome space, and external environment are reported. We consider the factors that
affect the development of microturbulence in the near-mirror air layer and inside the dome space,
variation of the telescope focal length with the temperature of its structures, variation of seeing
due to temperature gradients inside the primary mirror of the 6-m telescope. The methods used in
various observatories for reducing microturbulence are analyzed. We formulate suggestions concerning
the improvement of the temperature monitoring system currently in operation and the system of
automatic adjustment of the telescope focal length to compensate the thermal drift of the focus
during observations.

\end{abstract}
\maketitle

\section{Introduction}
\subsection[Mirror Thermal conditions and Seeing Degradation]{Effect of Telescope Mirror Thermal
Condition on Degradation of Astronomical Seeing}
The most commonly used method for the analysis of seeing involve the use of a differential image
motion monitor (DIMM). Such observations were performed in~1993 at Siding Spring
Observatory~\cite{1995PASA...12...95W}. The results were compared with the seeing achieved at
the~focus of the 3.9-m Anglo-Australian telescope (AAT). In addition to the estimates mentioned
above, this analysis used the temperatures of the primary mirror, dome space, and external
environment.

The above authors found seeing to be strictly anticorrelated with the temperature difference between
the mirror and the dome space in the case where the mirror temperature was higher than the external
air temperature: seeing deteriorated by~$1''$ compared to DIMM readings with each degree of
temperature difference. No seeing deterioration was observed in the cases where the mirror was
cooler than the dome space air.

A compact cooling facility has been developed to improve seeing in observations with the
\mbox{8.3-m}~Subaru telescope (Japanese instrument installed on Maua-Kea in Hawaii): the telescope
mirror was blown with cool air right inside the mount~\cite{2003PNAOJ...7...25M}. Such a location of
the cooling system allowed bringing the mirror to the operating condition much more efficiently.

Correlating the night temperatures based on weather forecasts for Mauna-Kea with
the actual night temperatures the above authors pointed out that the difference did not
exceed~$\pm2\degr$C in~80\% of the cases. Thus, given a weather forecast, it became possible to
bring the mirror to temperatures that would be certainly lower than the night temperature well in
advance (in daytime). To prevent the formation of condensate on the mirror and inside the mount, a
complex facility has been developed to control the temperature of the mirror and that of the cool
air in the
control system in order to avoid the dew point. This facility was also partially used during
observations: the compressor evacuating hot air from under the mirror continued to operate. As a
result, the air heated by the actuators of the active mirror was evacuated from the mount, thereby
reducing the heating of the Subaru mirror by the telescope mechanics.

Measurements showed that even such preobservational preparation of the mirror improves seeing by
at least~$0.1''$. Given the slower response of the primary mirror of the 6-m~telescope, introducing
a similar facility for it would bring even better results. However, in this case problems arise with
ensuring fast uniform cooling of the entire volume of the primary mirror of the 6-m~telescope and
with the accuracy of weather forecasts.

Experiments with a heated spherical mirror showed that laminar air flow with a velocity of
about~$1\ms$ can ``blow off'' microturbulent currents arising in the near-surface air
layer. Without active air blow, microturbulent currents in the near-mirror layer degrade seeing
at~$Z<45\degr$~\cite{1979MNRAS.188..249L}.

Seeing degradation is most pronounced at small mirror inclinations, i.e., for objects
with~$Z\lesssim10\degr$. For these areas seeing degradation is a factor of three to
five stronger than in the cases where the mirror is tilted by~$50\degr$. However, laminar airflow
blowing over the mirror surface at a speed of~$1\ms$ improves seeing significantly at small~$Z$ (at
large~$Z$ this airflow has the opposite effect degrading the seeing).

The best way to ``blow off'' microturbulent currents in the near-surface layer of large mirrors
would be to place the air blower horn in the central hole of the mirror for the air flow to spread
off the center.

An automatic system for producing the air flow to blow the surface of the mirror of the
6-m~telescope would make it possible to significantly improve the seeing for observations
at~$Z\lesssim10\degr$ even in the cases where the mirror temperature exceeds significantly
the ambient air temperature (by~$5\div10\degr$C). By \mbox{analogy} with the 2.5-m~INT telescope
(during its operation in the Greenwich Observatory), one can say that such measures would reduce the
seeing FWHM in the cases of maximum temperature differences for the 6-m~telescope (when the mirror
temperature exceeds the ambient air temperature by~$10\degr$C) down to about one and a half widths
of the turbulent atmospheric disk.

In order to develop active optics, a prototype telescope has been made at the National Astronomical
Observatory of Japan with a monolithic 61-cm~diameter 2.1-cm~thick primary pyrex
mirror~\cite{1991PASP..103..712I}. The shape of the mirror was controlled by 12~actuators.
A Shack--Hartmann sensor was used to analyze the wavefront. This experiment had a byproduct
consisting in the analysis of seeing variations caused by turbulence in the near-mirror air layer,
arising when the temperature of the mirror was higher than the air temperature inside the dome.

Wavefront distortions can be analytically characterized by the Zernike coefficients. As a combined
seeing characteristic, we can use the Strehl ratio:
$$
S = 1 - 2\pi^2 \mathrm{STR}^2 / \lambda^2,
$$
where $\mathrm{STR}^2$ is the weighted sum of squared Zernike coefficients over the radial order.
The closer is~$S$ to unity, the better will be the seeing.

Experiments performed by the above authors also showed seeing to degrade with increasing temperature
of the primary mirror because of turbulent air currents arising in the near-mirror layer. To ensure
seeing~$S>0.8$, the mirror temperature should be maintained at a level no higher than~$1\degr$C
above the ambient air temperature inside the dome.

To reduce the effect of microturbulence in the near-mirror layer, the above authors proposed to
``blow off'' microturbulent currents by laminar air flow. To this end, an air blower with a wide
horn was installed near the edge of the mirror. The laminar air flow blowing above the mirror
surface at a speed of~$1\ms$ (results identical to those obtained by~\cite{1979MNRAS.188..249L})
can reduce the seeing degradation by a factor of two (in terms of the Strehl ratio). For seeing
degradation to become measurable, it is sufficient to heat the mirror by~$0.2\degr$C above the
ambient air temperature.

Turbulence is usually characterized by the Rayleigh number. However, in our case a component of the
Rayleigh number, the Brunt--V\"ais\"al\"a frequency~\cite{1991PASP..103..712I}, can be used as an
objective characteristic of turbulence:
$$
N = \sqrt{
  -\dfrac{G}{\rho(0)}\partder{\rho(h)}{h}
},
$$
where $G$~is  the gravitational constant, and~$\rho(h)$~is the vertical density profile of gas.

Convective instability sets in if~$N^2 < 0$. We now obtain, passing from density to air temperature:
$$
N^2 = (G/c_p - \partder{T}{h})G/T,
$$
where $c_p$~is  the specific heat capacity at constant pressure, and~$T$~is the absolute air
temperature.  Thus, reconstructing an accurate vertical theoretical temperature profile in the air
column would allow is to determine whether air in it is convectively stable or
form turbulent currents.

The root mean squared velocity of turbulent currents can be computed from known air temperature
gradients:
$$
v^2 = (\grad T_h - g/c_p) Lg / \sqrt2 T,
$$
where $L$~is the mixing length (about~$5\text{--}20\,$cm).

However, serious technical challenges should be addressed to measure the parameters of
microturbulence arising inside the dome and in the near-mirror air layer by analyzing temperature
gradients: one has not only fill the dome space with many temperature sensors, placing them apart
from each other no farther than the mixing length, but also very accurately calibrate these sensors
and take readings frequently enough with an accuracy of several hundredths of a degree!

\subsection[In-dome conditions and seeing degradation]{Effect of Thermal Condition Inside the Dome
on Degradation of Astronomical Seeing}
Models of microturbulent air motions inside the dome are special cases of the Kolmogorov model for
closed finite volumes. The most popular among them are the von Karman, Greenwood--Tarazano, and
exponential models~\cite{2012SPIE.8444E..31Z}.

The refinement of a particular turbulence model is necessary for computing the characteristic size
of turbulent regions and the coherence time. One of the possible options for the microturbulence
detector is a device whose principle of operation is similar to that of the DIMM. In its base
configuration, the device has the form of a fiber laser beam splitter to produce several coherent
rays, which are transformed by a collimator into about 2-cm~diameter parallel light beams ($2\,$cm
is the minimum size of a microturbulent region). The beams pass the region studied at different
distances from each other and are then directed  by a system of diagonal mirrors to the focus of a
small telescope. Such a device can be used to study the characteristic spatial and temporal
frequencies of microturbulence.

Turbulent air flow can be generally characterized by the standard deviation of the velocity of air
currents, $\sigma_v$, or the fractional fluctuation of velocity,  $I = \sigma_v/\overline{v}$, where
$\overline{v}$~is the average velocity of air currents~\cite{1989Msngr..55...22Z}.
 
Empirical measurements of the turbulent current properties inside the dome can be performed with a
sensitive mechanical or ultrasound anemometer.

\subsection[Thermal Conditions and Primary Focus Position]{Effect of Thermal Condition of the
Primary Mirror and Its Metal Structures on the Position of the Focal Plane}
In the case of long-exposure photometric and spectroscopic observations (this especially concerns
photometry), it is very important not only to maintain the image at the same position but also to
maintain the constant focal distance. The focal distance of a telescope varies because of thermal
deformations of its supporting structures and optical surfaces.

Variations of the focal length of the telescope resulting from thermal deformations of the mirror
and supporting structures usually have the same sign (if only the operating surface of the mirror is
open and all other sides are thermally insulated), and therefore in the case of sufficiently slow
temperature variations the focal plane remains fixed relative to the flange of the primary focus.

If the mirror is not sufficiently thermally insulated, then thermal variations of the focus due to
mirror bending and variation of the length of the telescope rods would have different signs: in the
case of heating the curvature radius of the mirror decreases, whereas the rod length
increases~\cite{1918PASP...30...55P}.

Although thermal deformations of the mirror show up much slowly than the rod deformations, they also
contribute to the overall focus shift. Hence in the simplest linear model (if we neglect many other
factors causing the shift of the telescope focus) the focal length of the telescope depends on the
temperature of the primary mirror and the rod temperature as:
$$
F = F_0 + (T_M-T_0)\cdot f_M + (T_B-T_0)\cdot f_B,
$$
where $F_0$~is focus position at temperature~$T_0$;
$T_M$~is the temperature of the mirror;
$T_B$~is the temperature of the rods of the telescope ``tube'' structure;
$f_M$~is the linear contribution of mirror deformations to the variation of the focus;
$f_B$~is the linear contribution of the deformation of the supporting structures of the telescope to
the variation of the focus.

We thus see that even deriving the most elementary linear approximation to describe the temperature
dependence of the focus position requires constructing a two-dimensional experimental function: the
dependence of the focus on the mirror temperature and the temperature of the metal structures.
Because of the large mass of the 6-m~telescope primary mirror, thermal distortion of its shape is a
slow process, i.e., thermal variation of the position of the focus during the night would be
determined by the temperature of metal structures exclusively.

However, even such a simple approximation is inexact: the deformation of the mirror depends more
on the temperature gradient field inside the mirror rather than on the temperature proper. We are
thus dealing with a three-dimensional function even if we ignore less significant factors affecting
the position of the focus.

The results of theoretical calculations lead us to conclude that reconstruction of the exact
dependence of the telescope focal length on the temperatures of its constituent structures is a
rather complicated task. However, we can try to reconstruct with some accuracy a particular
dependence: the temperature dependence of the focal length during a certain observing night. The
temperature of the mirror cannot change significantly overnight, and therefore focus variation is
determined mostly by the contribution of the metal structures of the telescope. The error in this
case is determined by many external factors (including the variation of seeing during the night),
and therefore we cannot guarantee high accuracy even in terms of this approximation (we even cannot
guarantee that automatic correction of the position of the focal plane would be sufficiently
accurate to within~$\pm0.1\,$mm).

\section{Temperature Monitoring System for the Units of the 6-m~Telescope and the Air Inside the
Dome}
According to the documentation~\cite{BTAtech}, at the butt end of the primary mirror of the
6-m~telescope a total of sixty 310-mm~diameter 430-mm~deep blind holes have been made to accommodate
the mirror support unit (57~unloading and 3~defining support mechanisms). Also the mirror has six
about 10-mm~deep areas for jacking pads and one through hole with a diameter of~350\slash360~mm for
the centering shell.

The mirror is a~$655\pm0.3\,$mm~thick $6050\pm5\,$mm~diameter meniscus with surface radii of
$48050\pm50\,$mm. Each support mechanism shares~$1/60\,$part of the mirror weight (i.e.,
about~$700\,$kg). The upper part of the support is separated from the mirror by an air gap, where a
temperature sensor can be placed.

In December~2008 the technical specification for the development of a multichannel data acquisition
system to measure and store the temperatures of the 6-m~telescope units, dome, and interpanel space
has been approved. V.\,M.~Kravchenko was in charge of installing temperature sensors, and
S.\,I.~Sinyavskii dealt with connecting them to the data acquisition system.

Temperature measurements are made with uncalibrated TS-1288~temperature sensors with an accuracy
of~$0.5\degr$C or better. The M-FK-422~detectors were used to measure the temperatures of the
telescope structures and dome panels (the temperature measurement error does not
exceed~$0.5\degr$C). The detectors were not calibrated, and therefore their readings may differ by
more than~$0.5\degr$C.

Installation of sensors was completed by June~2010, however only part of them have been connected
to the data acquisition system. The sensors are connected to networked~TRM-138 eight-channel gauge
devices. Some of the sensors measure the air temperature, and some of them~--- the temperature of
metal structures and side surfaces of the 6-m~telescope primary mirror.

The data archiving system and the web interface~\cite{BTAthermal} for working with the archive were
developed by S.\,V.~Karpov. The readings from all temperature sensors connected to the system are
written to a PostgreSQL database at 15-min~intervals. The web interface or the local service allow
the required data to be extracted for any time interval when the system was operational.

Sensors for measuring the temperature of the primary mirror are arranged nonuniformly, and therefore
to measure the temperature gradients inside the mirror, these sensors should be not only calibrated
but also rearranged more uniformly. Placing temperature sensors in the hollows for support
mechanisms would make it possible to monitor temperature gradients inside the mirror body. The
linear thermal expansion coefficient, $\alpha_L$, of the S-316~glass used to make the blank
of the primary mirror of the 6-m~telescope is equal
to~$(3.0\div3.1)\cdot10^{-6}\,\degr\text{C}^{-1}$. For~$\alpha_L=3\cdot10^{-6}\,\degr\text{C}^{-1}$ 
a one-wavelength~($\lambda=500\,$nm) large warping of the shape of the mirror between two chosen
points would be observed in the case of the~$0.25\degr$C temperature difference (if the temperature
gradient is strictly radial). Thus, temperature registration for the 6-m telescope primary mirror
should be accurate to within at least~$\pm0.1\degr$C.

A proper choice of the type of temperature sensors for measuring the distribution of temperature
gradients in the 6-m telescope primary mirror would allow the calibrating temperature interval to be
reduced based on the accumulated statistics of mirror temperatures. For instance, the temperature of
the mirror remained within the~$-7\degr$C to~$+21\degr$C interval throughout the three-year interval
from December~23, 2010 through January~23, 2014. Such a reduction of the temperature interval makes
it possible to significantly improve the measurement accuracy.

In the temperature interval mentioned above, digital TSic~506 temperature sensors with the ZACwire
interface can be used. The operating temperature measurement interval of these sensors spans
from~$-10\degr$C to~$+60\degr$C. The rated accuracy throughout the entire operating range
is~$\pm0.3\degr$C or better (with a discretization of about~$\pm0.03\degr$C). According to
the rating data, a twofold reduction of the width of the measured temperature range allows the
relative accuracy of measurements to be significantly improved. Calibrating the sensors for the
temperature interval mentioned above reduced the measurement error down to~$\pm0.1\degr$C.

Another type of temperature sensors, TSYS~01 (with the SPI~and I${^2}$C interfaces), have a rated
accuracy of~$\pm(0.05\div0.1)\degr$C and the operating temperature range spanning fromÔ~$-40\degr$C
to~$+125\degr$C. The use of such digital temperature sensors would simplify the temperature
monitoring system and significantly improve its accuracy.

\section{Reduction of Temperature Data}
\subsection{Thermodynamics of the Dome Space}
The air volume in the dome space is equal to about~$23000\,\text{m}^3$.
Its variation with temperature is due mostly to the following processes:
\begin{enumerate}
\item\label{insolation} heating of the dome walls during daytime as a result of partial absorption
of sunlight (insolation);
\item heat exchange with the ambient environment via dome walls (mostly as a result of heat
transfer);
\item heat transfer as a result of the contact of air with the floor and lower part of the walls of
the dome space;
\item penetration of warm air currents from the lower levels of the tower into the airspace of the
dome room (oil exchange, space heating, and other systems).
\end{enumerate}
To protect the dome space of the 6-m~telescope against heating due to insolation, the dome has been
made with double-layered walls and ventilated gap between the walls. In addition, such a
double-layered structure of the dome walls reduces the efficiency of heat exchange with the ambient
environment.

According to the report by L.\,I.~Snezhko~\cite{1993SAOB...45...39S}, the qualification of the
thermal protection system of the 6-m~telescope performed in~1994 showed that:
\begin{itemize}
\item the dome efficiently protects the telescope against
insolation (i.e., the effect of the factor mentioned in
item no.~\ref{insolation}~above has been reduced practically to~zero);
\item  temperature inside the dome is distributed uniformly (to within the sensitivity of the method
of detection employed), only heating by~$0.2\degr$C of the
prime focus cell by observational equipment is detected;
\item the main source of heat release, the oil feed system, stands out by the high temperature of
oleoducts and elevated (by about~$\sim1\degr$C)  temperature of the slewing ring and exit points of
warm air from the hydrostatic bearing room.
\end{itemize}

In 2002 extensive work has been carried out in order to upgrade the oil feed system of the engines
of the 6-m~telescope (Yu.\,M.~Mametiev, A.\,M.~Pritychenko)~\cite{BTA_oil}. The upgrade reduced
substantially the power consumption and allowed cooling the oil before feeding it to the drives. As
a result, the heat output to the dome space of the 6-m~telescope has been reduced.

To cool the air in the dome space to maintain the temperature of the primary mirror of the
6-m~telescope in the working range, the conditioning system has been overhauled
(PishcheAgroStroyProekt, OAO). Three 30-kW freon compressor units were installed in the technical
unit located outside the tower of the 6-m~telescope. These compressor units are
used to cool the coolant consisting of 35\%~ethylene glycol solution, which is then pumped to air
exchange devices installed in the dome and under-floor space. To thaw the ice that forms on
air-exchange devices, electrical heaters with a total power of about 100\,kW are used.

The icing intensity on cool conductors and air interchanges increases with increasing temperature
difference between the dome space and coolant, resulting in increased energy consumption for
defrosting and reduced efficiency of the upgraded cooling system~\cite{BTA_cold}.

The efficiency of this system can be demonstrated by simple thermodynamic computations.

\subsubsection{Adiabatic Approximation}
We first compute the time required for the cooling system to reduce the air temperature in the dome
space by~$10\degr$C (to achieve the maximum permissible mirror cooling rate, $2\degr$C per
day~\cite{1993SAOB...45...39S}).

In the absence of heat exchange with the ambient environment air temperature varies isochorically
(with a slight change of pressure due to the change of temperature), then isobarically (due to the
air exchange with the ambient environment). Strictly speaking, isobaric variation of temperature
alone is sufficient to compromise the adiabatic approximation considered, and therefore we assume
the process to be isochoric to a first approximation.

Let us now compute the amount of heat to be withdrawn from the dome space air to cool it
by~$dT\degr$C (hereafter we adopt the main computational
formulas from~\cite{Sivuh, Savel}):
$$c_V=\frac{i}2\frac{R}{M},\qquad
dQ=mc_V\,dT=\rho V\frac{i}2\frac{R}{M}\,dT$$
$$dQ=1.29\cdot23000\cdot\frac52\frac{8.31}{29\cdot10^{-3}}\,dT=2.13\cdot
10^7\,dT\;[\text{J}],$$
where $c_V$~is the constant volume specific heat;
$i$~is the number of degrees of freedom of ideal-gas particles;
$M$,~the molar mass of the air;
$R$,~the universal gas constant;
$\rho$,~the air density;
$Q$,~the amount of heat, and
$T$,~the temperature.

Thus to cool the dome space air by~$10\degr$C, we have to withdraw from it about 213\,MJ of energy.

The total cooling capacity of the cooling system is of about 107\,kW. The minimum time required to
withdraw 213\,MJ of energy from the dome space air (without the time required for forced mixing of
air) is in this case
$$t_{min}=\frac{2.13\cdot10^8}{1.07\cdot10^5}\approx2000\,\text{s},$$
i.e., about 33\,min.

\subsubsection{Computation of Heat Inflow}
The heat current intensity from the floor and walls of the dome space depends on temperature
gradient:
$$\frac{\partial Q}{dt\,dS}=-\kappa\partder{T}{x},$$
which increases sharply in the case of intense mixing (here $\kappa$~is the air heat-conduction
coefficient). Thus the heat output for the given gradient $\theta=\grad T$~is
$$P=-24.1\cdot10^{-3}S\theta\;[\text{W}],$$
where $S$~is the area of the surface with the given gradient~$\theta$. For example, the heat output
of the floor is
$$P=-13.7\theta\;[\text{W}].$$
The heat output of the dome walls can be computed in a similar way ($P\approx-60\theta\;$).

Such a small heat current allows us to neglect the heating of air inside the dome caused by heat
transfer from the walls and the floor. Hence the main source of air heating inside the dome are heat
currents from living quarters and technological buildings.

L.\,I.~Snezhko approximately computed the heat currents, which proved to correspond to a temperature
increase rate of~$1.4\degr\text{C\,h}^{-1}$ equivalent to a heat output of about 8.3\,kW.

Approximate computations show that the available cooling system, if it were perfect, would be
capable of maintaining the temperature inside the dome at a level corresponding to the temperature
difference of up to~$10\degr$C between the primary mirror and dome space air. The cooling capacity
required to maintain the air temperature inside the dome at a given level can be estimated at
9--10\,kW (L.\,I.~Snezhko). However, given that the capacity of the air conditioning
system inside the dome of the 6-m~telescope depends heavily on the external air temperature and the
difference between the external and internal temperatures, the efficiency of this system in
wintertime would be low.

To achieve the maximum permissible cooling rate of the primary mirror, the air inside the dome
should be gradually cooled until the mirror reaches the required temperature while maintaining
constant temperature difference between the mirror and air in the process.

A detailed temperature monitoring of the dome space is needed to more accurately determine the
temperature gradients and heat output of stray heat sources (and hence the amount of electric power
needed to maintain the given temperature in the dome space).

\subsection{Focus Position Temperature Dependence}
\begin{pict}
\includegraphics[width=\columnwidth]{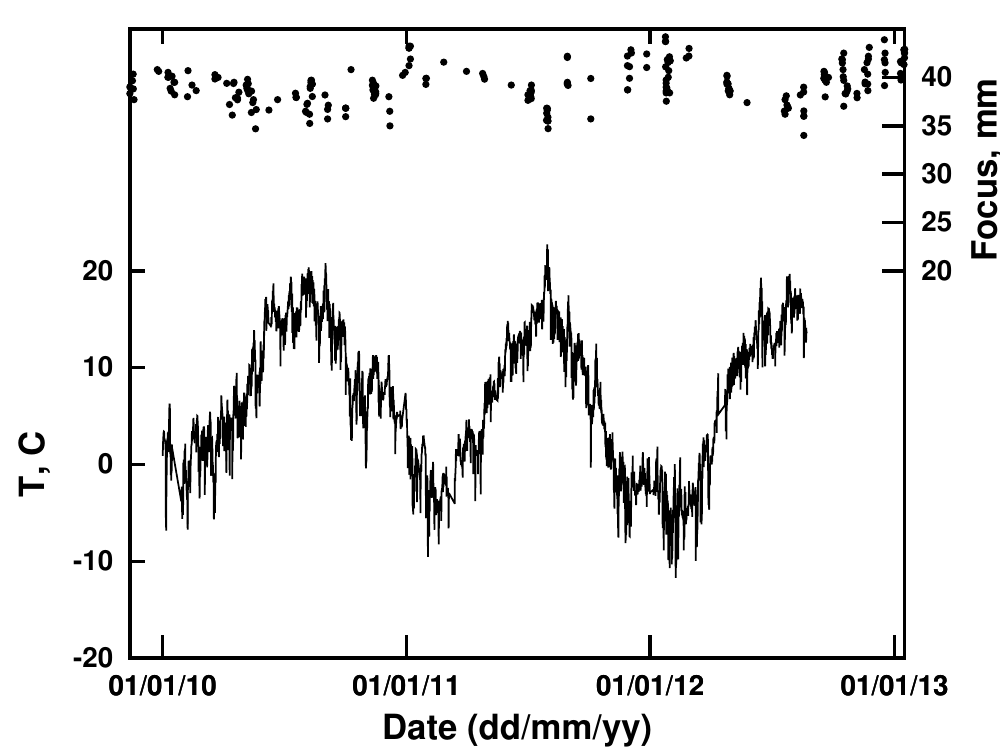}
\caption{Rod temperature (the solid line) and focus readings (the dots) over three years.}
\label{tempfoc3yr}
\end{pict}

To analyze the dependence of the position of the
focus on the temperature of different telescope and
dome parts, we selected the appropriate data for the
past three years from the temperature archive and
from ASPID database~\cite{BTA_aspid}. To this end, we
\begin{enumerate}
\item\label{filenames} recursively scanned the directories with the archives to select from SCORPIO
logs the strings following the text ``focussing TELESCOPE'' (i.e., information about images acquired
after focussing the telescope);
\item generated archive names based on the names of retrieved files they contain;
\item unpacked all archives that we found to be of interest for us into the current directory;
\item copied the FITS header of the files whose names we obtained at step~\ref{filenames}
into a separate file;
\item analyzed the \verb'INSTRUME' and \verb'IMAGETYP' fields of the FITS header to determine
whether the given image was acquired on SCORPIO and whether the image type is ``object''; if the
condition was satisfied, we wrote into the log file the observing date and time, focus, and the
temperatures of the mirror, domespace, and external environment.
\end{enumerate}
\begin{pict}
\includegraphics[width=\columnwidth]{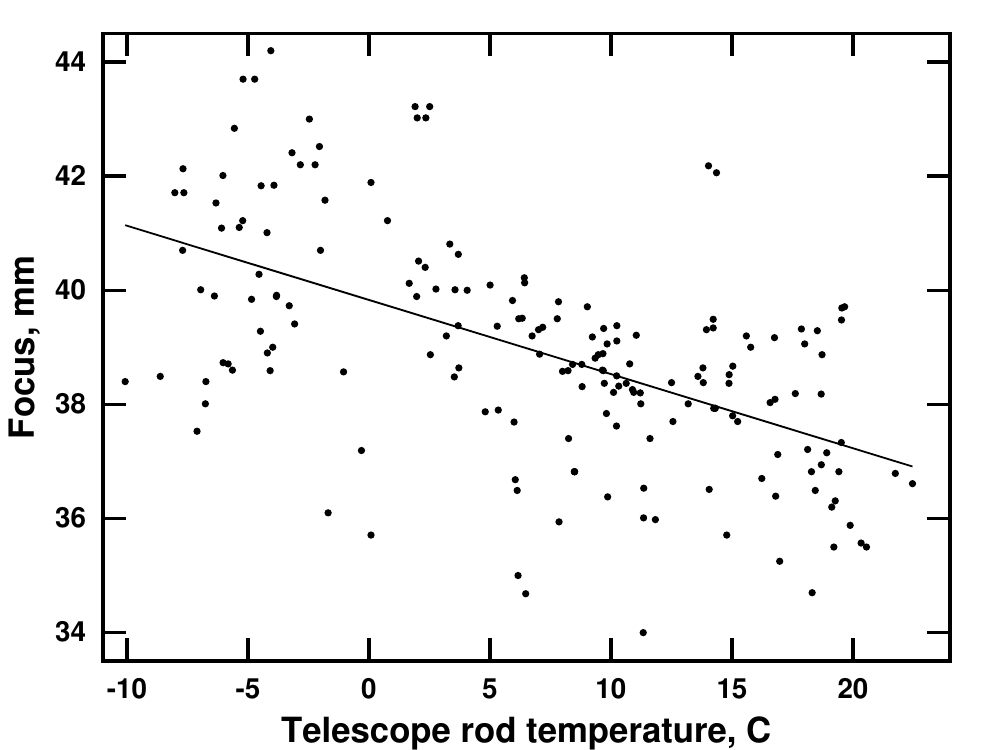}
\caption{Relationship between focus measurements and rod temperature. The solid line shows a linear
approximation.}
\label{FvsTsup}
\end{pict}
We retrieved from the temperature archive the readings of the thermometer located on one of the
telescope rods. We then averaged these data into 15-min~bins (the database contains readings made at
10-s~intervals). We performed all subsequent computations in Octave environment.

We adopted the focal lengths and temperatures from different archives based on measurements made
with different sampling times~\look{tempfoc3yr}, and therefore we had first to homogenize the
observational data reducing them to a single system. To this end, we selected those focus
measurements for which rod temperature measurements were available within 15\,min from the
focus-length measurement. We then interpolated the rod temperature to the image acquisition times on
SCORPIO.

The correlation coefficient between the rod temperature and the measured focus lengths is equal
to~$-0.56$, and the linear interpolation of the relation~\look{FvsTsup} yields the following
approximate formula
\begin{equation}
F \approx 39.83 - 0.13\cdot T,\qquad\sigma_F=1.64,
\label{FvsSupT}
\end{equation}
where $F$~are the focal length measurements in mm; $T$, the rod temperature in Celcius degrees, and 
$\sigma_F$, the standard deviation of the focal length.

\begin{pict}
\includegraphics[width=\columnwidth]{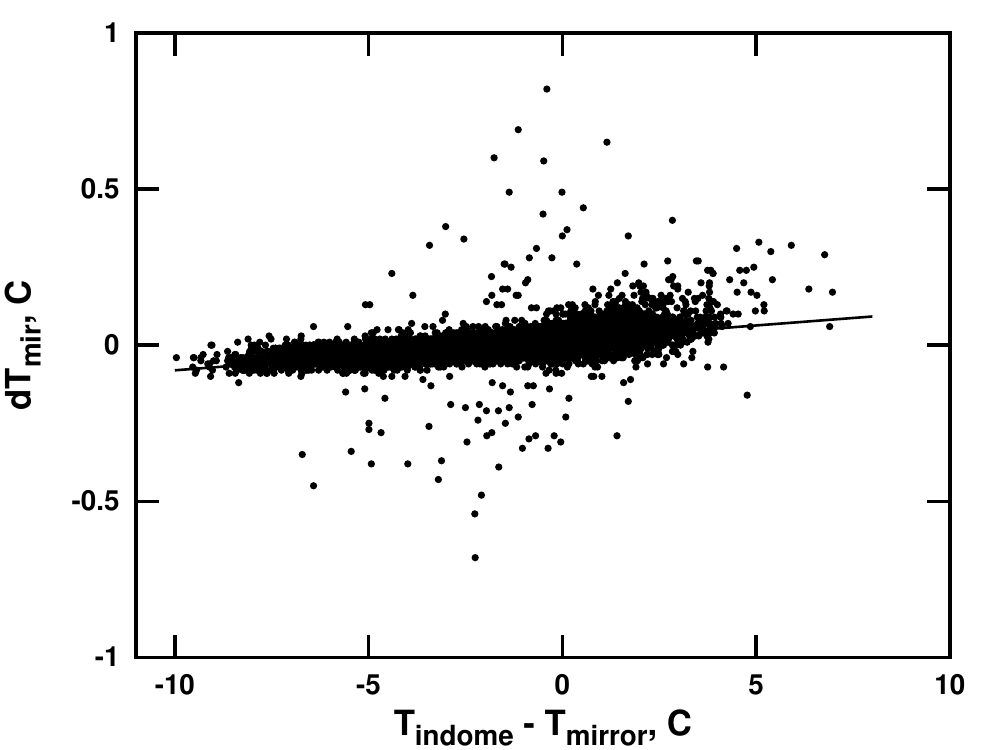}
\caption{Relation between the rate of mirror temperature change and the temperature difference
between the dome space and the mirror (data for the three-year period). The solid line shows a
linear fit to the data.}
\label{dTM_vs_dT}
\end{pict}

An analysis of similar relationships between focus measurements and dome space and mirror
temperatures yielded the following results. The correlation coefficient between the dome space
temperature and focus measurements is equal to~$-0.54$, and the corresponding linear approximation
has the form $F \approx 39.53 - 0.13\cdot T$, $\sigma_F=1.67$, which is close to~\eqref{FvsSupT}.
The correlation coefficient for the mirror temperature was the highest among all ($-0.66$), and
the corresponding linear approximation has the form $F\approx 40.22 - 0.17\cdot T$, $\sigma_F=1.49$.
It is clear that in the latter case the resulting relation is more well defined, because the mirror
acts as sort of temperature damper, which smooths rapid variations affecting the general
correlation.

A detailed analysis of individual nights did not improve the statistics: the scatter of the data
about linear approximation still remained too large to conclude about bona fide unique dependence of
focus measurements on the readings of any temperature sensor.

The large data scatter in our case is due not solely to various errors, but also to the more complex
form of the temperature dependence of the telescope focal length. Thus, e.g., in the case the
detector is held by a sole steel rod and there are no other factors involved, the coefficient in
dependence~\eqref{FvsSupT} would be equal to~$-0.31$. Temperature gradients arising inside the
mirror body in response to a change of its temperature would give rise to another, positive
coefficient, and a well-defined dependence on the rate of change of the dome space temperature
(because the variation of the temperature of the primary mirror of the 6-m~telescope also depends on
on it).

\begin{pict}
\includegraphics[width=\columnwidth]{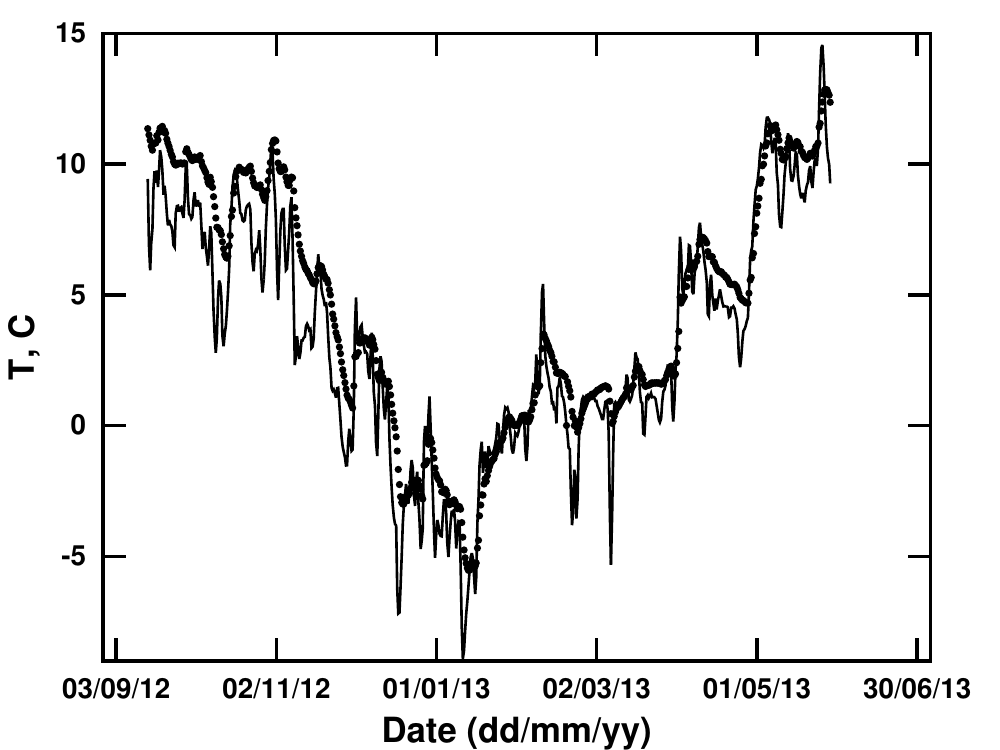}
\caption{Temperature of the mirror (the dots) and one-day moving average of the dome-space
temperature (the solid line) for a selected time interval.}
\label{mdtemps}
\end{pict}

We can thus conclude that unambiguous automatic adjustment of the focal length in the process of
observations without the need to readjust the focus by observing stars can be best achieved by
installing a laser range finder on the prime focus cell and monitoring the distance from the prime
focus cell to the spider support of the diagonal mirror or some point on the mirror mount. However,
a careful calibration of temperature sensors and their rearrangement will make it possible to derive
a unique analytical dependence.

\subsection[Dependence on the in-Dome Temperature]{Dependence of the Rate of Primary Mirror
Temperature Change on the in-Dome Temperature}
To construct this dependence, we retrieved from the archives the data concerning the temperature
inside the dome and the temperature of the mirror averaged in one-hour bins. After differentiating
the mirror temperature, $dT\ind{mir}$, and computing the temperature difference $\Delta T$ between
the dome space and the mirror, the resulting datasets were cleaned to remove analog-to-digital
converter errors and surges due to the nonuniform distribution of measurements by rejecting the
measurements with~$|\Delta T| > 10\degr$C and $|dT\ind{mir}|>1\degr$C.

The correlation coefficient between $dT\ind{mir}$ and $\Delta T$ is~$0.52$. Linear fit to all the
data obtained~\look{dTM_vs_dT} has the form:
$$
dT\ind{mir} \approx 0.0152 + 0.0096\cdot\Delta T.
$$

\begin{widepict}
\includegraphics[width=0.48\textwidth]{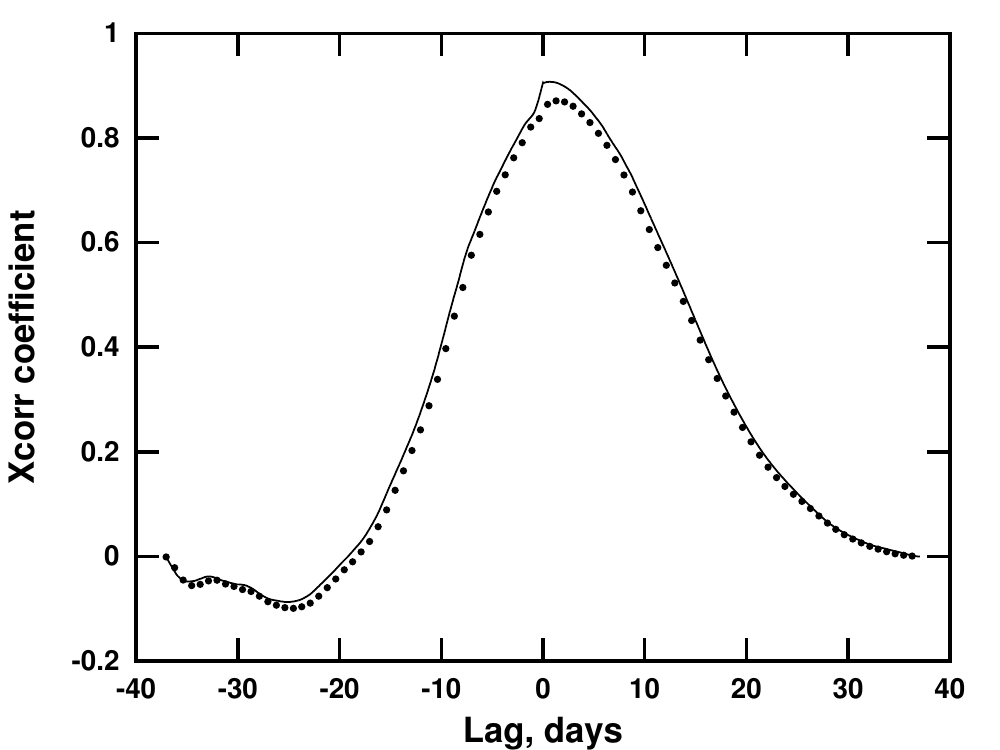}\hfill
\includegraphics[width=0.48\textwidth]{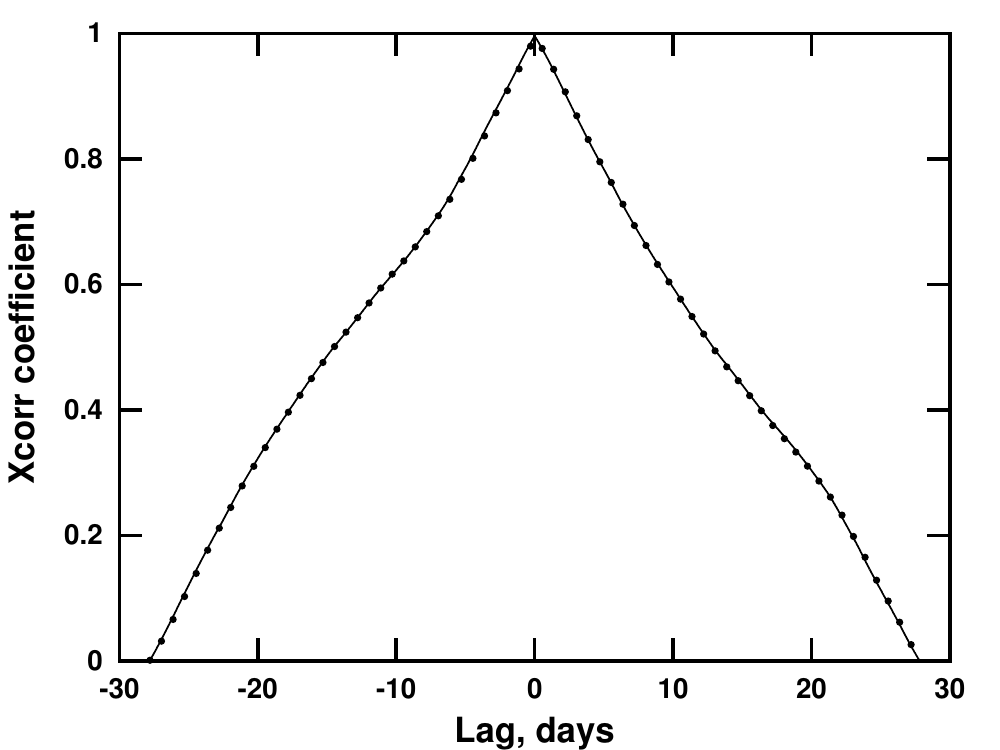}
\caption{Cross-correlation function of the temperature of the 6-m telescope primary mirror and that
of the dome space. The dots and solid lines show the temperatures averaged in one-hour and one-day
bins respectively.}
\label{mdtempscorr}
\end{widepict}

To perform the correlation analysis, we selected from the resulting data set ten time intervals
with uniformly distributed temperature measurements. The longest continuous interval~\look{mdtemps}
spanned about 255~days (from September~14, 2012 through April~28, 2013). For each interval we
constructedthe cross-correlation functions between the temperatures of the mirror and dome space,
and derived linear approximations between the rate o change of the mirror temperature and that of
the temperature difference between the dome space and the mirror.

The position of the maximum of the cross-correlation function for all selected time intervals varied
from 41~hours (31~hours for the moving average) for January~2011 to zero for the other our
intervals. The largest shift evidently is obtained for the case of time intervals with sharp changes
of the temperature of the dome space. Zero shift corresponds to time intervals with smoothly varying
daily averaged temperature, and also time intervals when the air conditioning system was switched on
in the dome. For example, Fig.~\ref{mdtempscorr} shows the mutual correlation functions of the
mirror and dome space temperature for the cases of nonzero and zero shifts.

Fitting the rate of change of the temperature of the primary mirror of the 6-m~telescope as a
function of the temperature difference between the dome space and the mirror for all ten intervals
yielded an average coefficient of~$0.01\,\text{h}^{-1}$, which agrees with known values: for the
temperature of the primary mirror of the 6-m~telescope to change by~$0.1\degr$C in one hour, the
temperature in the dome space should differ from the mirror temperature by~$10\degr$C.

The data obtained fully agree with the results of the computations performed by L.\,I.~Snezhko. To
study the processes caused by temperature gradients inside the primary mirror of the 6-m~telescope,
we have to place accurately calibrated temperature sensors in the supporting holes.

\section{Conclusions}
To sum up, we conclude that: first, there is no need to install a large number of temperature
sensors inside the dome; second, for unambiguous assessment of the extra degradation of seeing due
to temperature gradients and microturbulent air motions, it is necessary to measure the temperature
of telescope rods at several points and the distribution of temperature inside the primary mirror.
It is very important to ensure that all sensors have zero scatter of readings (within the
permissible error) in the temperature interval from~$-10\degr$C to~$+25\degr$C. This means that all
sensors should be accurately calibrated in a thermostat using a certified precision reference
thermometer. The sensor readings should be converted into real temperatures both inside the data
acquisition system and in the archiving server when the new portion of temperature measurements is
written to the database.

The use of digital temperature sensors would make it possible to simplify the temperature monitoring
system and improve its accuracy. Another option for the upgrade consists in the use of the available
system carefully calibrating and rearranging the temperature sensors.

\section{Acknowledgments}
I am grateful to Prof. V.\,L.~Afanasiev for suggesting the idea of analyzing the dependence of the
6-m~telescope focus position on temperature conditions and for describing the ASPID database
structure. I am also grateful to Prof. V.\,E.~Panchuk for valuable comments concerning the text of
the paper. Observations on the 6-m~telescope of the Special Astrophysical Observatory are held with
financial support from the Ministry of Education and Science of the Russian Federation (contract No.
14.619.21.0004, project identifier RFMEFI61914X0004).

\end{document}